# Tunable frequency conversion in doped photonic crystal fiber pumped near degeneracy

LEAH R MURPHY[1,*], MATEUSZ J OLSZEWSKI[1], PETROS ANDROVITSANEAS [1,2], MIGUEL ALVAREZ PEREZ[2], WILL A M SMITH [1], ANTHONY J BENNETT [2], PETER J MOSLEY [1], AND ALEX O C DAVIS [1]

[1]*Centre for Photonics and Photonic Materials, Department of Physics, University of Bath, BA2 7AY, United Kingdom*
[2]*School of Engineering, Cardiff University, Queen's Buildings, The Parade, Cardiff CF24 3AA, United Kingdom*
*\*lrm60@bath.ac.uk*

**Abstract:** Future quantum networks will rely on the ability to coherently transfer optically encoded quantum information between different wavelength bands. Bragg-scattering four-wave mixing in optical fiber is a promising route to achieving this, but requires fibers with precise dispersion control and broadband transmission at signal, target and pump wavelengths. Here we introduce a photonic crystal fiber with a germanium-doped core featuring group velocity matching at 1550 nm, the telecoms C-band, and 920 nm, within the emission range of efficient single photon sources based on InAs quantum dots. With low chromatic walk-off and good optical guidance even at long wavelengths, large lengths of this fiber are used to achieve nanometer-scale frequency shifts between wavelengths around 920 nm with up to 79.4% internal conversion efficiency, allowing dissimilar InAs dots to be interfaced. We also show how cascading this frequency conversion can be used to generate a frequency comb away from telecoms wavelengths. Finally, we use the fiber to demonstrate tunable frequency conversion of weak classical signals around 918 nm to the telecoms C-band.

## 1. Introduction and background

Frequency conversion of optical signals is an area of intense research interest, with applications including quantum technologies [1], nonlinear interferometry [2], frequency combs [3], and novel laser sources. In large-scale quantum networks, coherent frequency conversion is needed to interface various components such as photon sources, transmission lines, memories and processing nodes, which couple to diverse wavelengths [4]. For example, minimum attenuation in solid silica optical fiber is around 1550 nm, while efficient single photon sources based on InAs quantum dots (QDs) operate in the range of 900-950 nm [5, 6]. In addition, since high indistinguishability is required for most quantum information technologies, and any two dots seldom emit at precisely the same wavelengths [7, 8], frequency conversion between the emission wavelengths of dissimilar dots is also of interest for quantum information processing.

The photon statistics and coherence of an optical field have been shown to be preserved after frequency conversion by electro-optic modulation, which is suitable for sub-nanometer shifts of ultrafast pulses [9], as well as parametric nonlinear optical techniques such as sum/difference frequency generation (S/DFG), a $\chi^{(2)}$ process [10–13], and Bragg scattering four-wave mixing (FWM), a $\chi^{(3)}$ process [14–16]. S/DFG, which features a single pump whose wavelength is fixed by the signal and target via energy conservation, is effective when large frequency shifts are needed. Smaller shifts are inaccessible as they require either impractically short or long pump wavelengths. FWM instead involves photons at the signal (*s*) wavelength being translated to a target (*t*) wavelength by a frequency shift determined by the difference between two pump fields *p* and *q* (see Fig.1(a)). It is attractive for network applications as the interaction takes place inside optical fiber, which greatly simplifies low-loss integration with long-range fiber connections. An additional advantage of FWM is the flexibility engendered by having two pumps, which grants a free parameter in the choice of their wavelengths and permits small as well as large frequency shifts [17]. This enables a near-degenerate regime not accessible by S/DFG, where the FWM interaction couples two pairs of relatively nearby wavelengths (see Fig.1a). The same fiber can thus be used to achieve either small or large frequency shifts depending on the choice of pumps.

Here, we explore this regime using germanium-doped photonic crystal fiber (Ge-PCF). We show how working close to degeneracy offers many advantages including broad phase matching, long interaction lengths, and the proximity of both pump wavelengths within the gain region of a single optical amplifier. Firstly, we demonstrate tunable, efficient frequency conversion over several nanometers for picosecond pulses at various wavelengths from 915-922 nm— suitable for interfacing dissimilar InAs QDs. We thereby also demonstrate the feasibility of using a single commercially available fiber amplifier, seeded with a two-tone signal in the telecoms C-band, to provide both of the required pump fields. This considerably reduces the experimental cost and complexity of FWM-based frequency conversion. Secondly we show how cascading this frequency conversion can be used to generate a frequency comb around 918 nm, far from the wavelengths of the amplified pump fields. Lastly, we show the reversibility of the process, demonstrating tunable downconversion of pulses from the InAs emission band to the telecoms C-band and paving the way to a fiber-integrated true single-photon source at telecoms wavelengths.

## 2. Fiber design

Frequency conversion by FWM requires the simultaneous satisfaction of energy conservation ($\omega_p - \omega_q = \omega_t - \omega_s \equiv \Delta\omega$) and phase matching ($\beta_p - \beta_q = \beta_t - \beta_s \equiv \Delta\beta$), where $\omega_i$ and $\beta_i$ are the angular frequency and propagation constant of the field $i$, respectively [18]. In a refractive medium such as optical fiber, $\beta(\omega)$ is determined by the dispersion so FWM can generally only take place for certain combinations of wavelengths controlled by the fiber design.

The near-degenerate condition is characterized by small $\Delta\omega$, and hence we can expand

$$\beta_t \approx \beta_s + \left.\frac{\partial\beta}{\partial\omega}\right|_s \Delta\omega \qquad (1)$$

$$\beta_p \approx \beta_q + \left.\frac{\partial\beta}{\partial\omega}\right|_q \Delta\omega. \qquad (2)$$

The phase matching condition can therefore be written as

$$\Delta\beta = v_g(\omega_s)\Delta\omega = v_g(\omega_q)\Delta\omega, \qquad (3)$$

where $v_g \equiv \frac{\partial\beta}{\partial\omega}$ is the group velocity. Phase matching therefore reduces to matching the group velocities in the neighbourhood of both pairs of wavelengths. The difference in group velocity also determines the propagation distance at which significant phase mismatch accrues between wavelengths detuned from optimal phase matching, known as the walk-off length [18]. Typically, this introduces a limiting trade-off between phase matching bandwidths and available interaction length. Since group velocities are matched in the near-degenerate condition, here phase matching can be sustained over longer sections of fiber and larger bandwidths compared to more widely spaced wavelength schemes.

PCF is a type of microstructured fiber consisting of longitudinal air holes arranged periodically in a glass matrix, with a core region provided by a missing hole [19]. Guidance is provided by the index contrast between the all-glass core and the air-infiltrated cladding region. Greater index contrast than that present in conventional fiber ensures tight confinement, small effective mode areas and hence high nonlinearities. In addition, there is considerable flexibility in the design of PCF, with structures characterized by the diameter ($d$) and spacing ($\Lambda$) of the air holes, as well as the number of rings of air holes. This flexibility enables wide control over the dispersion of the waveguide which, combined with high nonlinearity, makes PCF a natural platform for

FWM technologies. It has previously been used to demonstrate quantum frequency conversion by FWM [20, 21].

Achieving the near-degeneracy condition, eq. 3, for wavelength pairs in the 920 nm range and the C-band involves fixing the dispersion curve of the PCF such that $v_g(\lambda_s) = v_g(\lambda_q)$, where in our case $\lambda_s \approx 920$ nm and $\lambda_q \approx 1550$ nm. The wavelength where $v_g(\lambda)$ is maximal, corresponding to the zero dispersion wavelength (ZDW), must therefore fall between the two (see Fig.1(d), $\lambda_{ZDW} \approx 1180$ nm). In practice, this calls for a relatively small modification of the ZDW from that of conventional telecoms fiber, which, due to the weakness of the waveguiding, is close to that of bulk silica at around 1300 nm. A suitable subspace of values for $d$ and $\Lambda$ for a pure-silica PCF was determined using empirical relations [22], finding that this phase matching condition calls for relatively small, closely spaced air holes. However, numerical modelling of pure-silica PCF indicates that fibers with these parameters experience high confinement losses at near infra-red wavelengths even when designed with many rings of holes [23], with losses in the C-band becoming unacceptably poor for our application. To overcome this problem, we modified our fiber design with the inclusion of a germanium-doped core region, which provides additional confinement by enhancing the index contrast between the core and the cladding while still allowing for dispersion control through the air hole parameters. This core region has a diameter of $0.74\Lambda$, with an approximately parabolic refractive index profile ranging from $\Delta n = 0$ at the edge to $\Delta n = 0.031$ at the centre (as measured relative to pure silica). The Ge-PCF structure was modelled by finite-element analysis in COMSOL Multiphysics, where a design with $\Lambda = 2.46$ μm and $d/\Lambda = 0.36$ was found to provide the required dispersion.

## 3. Fabrication and characterization

The Ge-PCF was produced at the University of Bath's fiber fabrication facility by the two-stage stack-and-draw technique [24], with the hand-assembled stack first drawn to millimeter-scale precursor canes before being drawn again to fiber. A micrograph of the Ge-doped cane is shown in Fig.1(b), with the approximately hexagonal doped core region clearly visible. Several sections of fiber were produced with the target pitch, while the pressure used to inflate the air holes was varied between sections to sweep the hole diameter over the target value. We then characterized the dispersion of these fibers by spectral interferometry [25]. The characterization setup is shown in Fig.1(c). Broadband supercontinuum pulses were produced by pumping Bath-made PCF with pulses from a 1064 nm microchip laser (Teem Photonics MNP-08E-100). These pulses were then routed into a Mach-Zehnder interferometer, with one arm containing a section of the Ge-PCF under test of length $L$ (typically 50–70 cm) and the reference arm containing a variable delay in free space. One output of the interferometer was resolved on an optical spectrum analyser (OSA, Yokogawa AQ6374). This revealed a spectral interference pattern, with fringe density falling to zero at the wavelength where the group delay in the two arms is equal. By varying the mirror position in the delay line, we mapped the variation in the group delay of the test fibers, $v_g(\lambda)L$, and hence determined $v_g(\lambda)$ up to a constant factor. (An absolute determination of $v_g(\lambda)$, which would have required a precise measurement of the free-space path difference in the interferometer, was not necessary). We thus arrived at a fiber for which the group velocity was matched for the InAs dot emission range and the telecoms C-band (see Fig.1(d)).

We characterized the attenuation of the Ge-PCF by performing a cutback from 10 m to 4 m, where the fiber was wound in loose loops of around 60 cm bend diameter to minimize bend loss. We launched light from a broadband Energetiq EQ-99X laser-driven light source into a short length of smf-28e which was butt-coupled to the Ge-PCF. We collected spectra using a Bentham DTMc300 double monochromator through a bare fiber adaptor. The attenuation was measured to be up to $0.03$ dB m$^{-1}$ in the InAs dot emission range and up to $0.05$ dB m$^{-1}$ in the C-band.

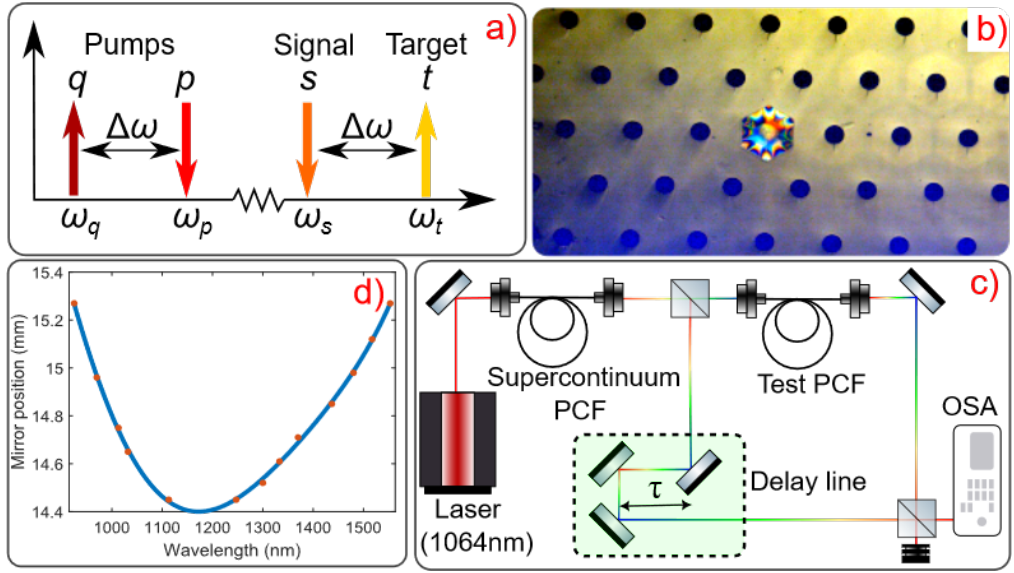

Fig. 1. (a) Wavelength scheme for near-degenerate four wave mixing, pumped for $s \to t$ upconversion. Photons are annihilated at pump frequency $\omega_p$ and signal frequency $\omega_s$, and created at pump frequency $\omega_q$ and target frequency $\omega_t$. (b) Optical micrograph of the cross-section of the cane used to draw Ge-doped fiber showing germania-doped core region. Lattice spacing is 160 µm. (c) Experimental scheme for fiber dispersion characterization. Broadband supercontinuum light from a PCF source is used to measure fiber dispersion by spectral interferometry in a Mach-Zehnder interferometer (MZI) with variable delay $\tau$. OSA- optical spectrum analyser. (d) Red points: wavelength dependence of the delay $c\tau$ needed to balance the MZI for a 63.5 cm length of fiber, with $\tau \propto v_g(\lambda)$. Blue curve: fifth-order polynomial fit. The minimum corresponds to the zero dispersion wavelength.

## 4. Experimental methodology

### 4.1. Upconversion within the QD-band

In the first phase of our experimental demonstration, two nanometer-detuned pumps in the telecoms C-band were used to drive frequency conversion of pulses within the emission range of InAs QDs. A schematic diagram illustrating the experimental setup is provided in Fig. 2. An 80 MHz repetition rate Ti:Sapphire laser (SpectraPhysics Tsunami) operating mode locked at various wavelengths between 914.5 and 921.5 nm (see Table 1) with a pulse duration of ~100 ps served as the signal source. The polarization and intensity of this laser was controlled using half-wave plates (HWPs) and a polarising beamsplitter (PBS) to control beams directed to a spectrometer, autocorrelator, and photodiode for calibration and monitoring purposes (not shown in figure).

The Ti:sapphire pump pulses were split at a PBS, with the reflection arm coupled into fiber and directed to an amplified Si photodetector, which provided an electronic signal used to carve nanosecond pulses in the C-band pump fields. The transmitted path directs ~98% of light from the Ti:sapphire laser towards the Ge-PCF to serve as the signal field, $s$, passing through a quarter waveplate (QWP) to reduce back reflections on the Ti:Sapphire. A Keplerian telescope was used to optimize the coupling efficiency of the Ti:Sapphire emission into the Ge-PCF, compensating for the chromatic dependence of the focal length of the fiber coupling lens. The C-band pump branch consisted of two continuous-wave lasers, one (CoBrite-DX1) operating either at 1555.7 nm or 1554.7 nm, and the other (Laser2000 TFL-170-1560-FCAPC) temperature-tunable between 1551.4–1560.9 nm. These fields were combined via a 50:50 fiber beamsplitter (FBS) and directed via polarization maintaining (PM) fiber to an electro-optic amplitude modulator (EOM) to undergo pulse carving synchronized to the Ti:sapphire laser, and amplified by an erbium-doped fiber amplifier (EDFA, BKTel HPOA-1560-S370A). At the output of the EDFA total power of the combined output was 3.94 W.

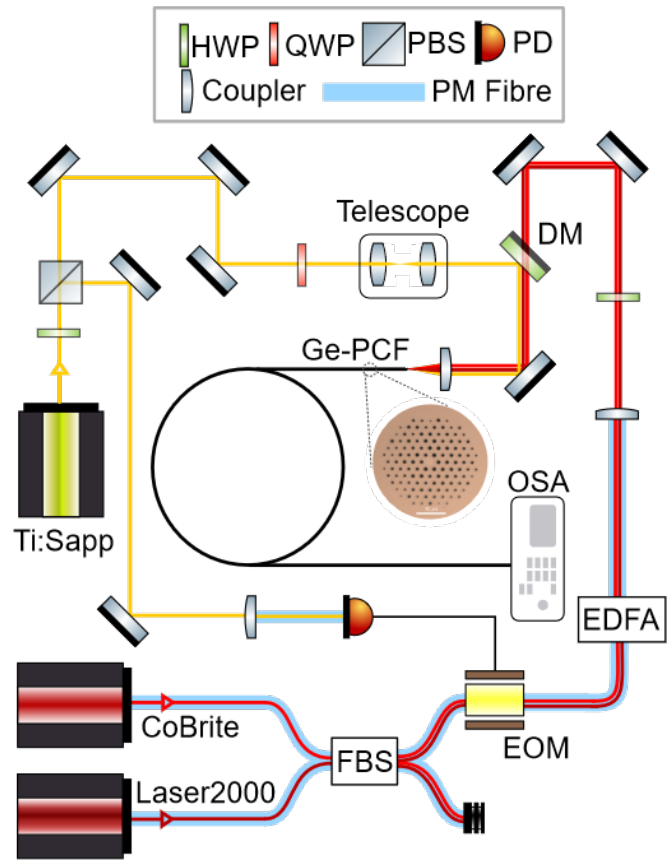

Fig. 2. Schematic showing the experimental setup for QD-band frequency conversion. Two C-band seed lasers are combined at a fiber beam splitter (FBS) before undergoing pulse carving at an electro-optic modulator (EOM) and amplification by an erbium-doped fiber amplifier (EDFA). These pump pulses are combined at a dichroic mirror with signal pulses from a Ti:Sapphire laser system and coupled into the Ge-PCF (cross-section shown). The output is monitored by an optical spectrum analyzer (OSA).

The C-band pump fields were combined with the signal field at a dichroic mirror (DM) before all three were directed into the Ge-PCF. An aspheric lens maximized coupling efficiency into the PCF across all three fields. A 34 m section of PCF fiber was used. The tunability of all three laser sources $\lambda_s$, $\lambda_p$ and $\lambda_q$ allows for a wide exploration of the wavelength dependence of the FWM process. The output end of the Ge-PCF was connected either to the optical spectrum analyzer (OSA) or a power meter for characterization.

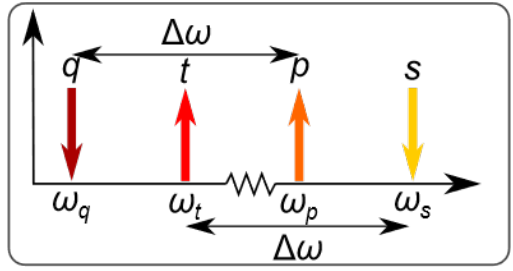

Fig. 3. Wavelength scheme for near-degenerate four wave mixing, pumped for $s \rightarrow t$ downconversion with a large frequency shift. Photons are annihilated at pump frequency $\omega_q$ and signal frequency $\omega_s$, and created at pump frequency $\omega_p$ and target frequency $\omega_t$.

### 4.2. Downconversion of QD-band pulses to the C-band

In the second phase of our experiment, two pumps (one in the telecoms C-band, $q$, and one in the InAs QD emission range, $p$) were used to downconvert a weak classical signal in the InAs emission range to the telecoms C-band (see Fig. 3). The C-band pump was prepared similarly to above. To emulate more faithfully a single photon emitted from a QD, the signal, $s$, was provided by a 3 ps pulse from a Ti:Sapphire mode-locked laser (MIRA 900P) stretched to $\sim$ 60 ps by spectral filtering with the use of a $\sim$ 1200 lines/mm transmission grating, with a fiber-coupled power of 3 mW. For the QD-band pump, a CW field was used from a Ti:Sapphire laser ($M^2$ SolsTiS) with a fiber-coupled power of 20 mW. All lasers were combined with the same linear polarization. The signal and the QD-band pump were collimated using two aspheric lenses with a focal length of 8 mm and combined using a bandpass filter with a bandwidth of 200 pm and central wavelength of 935 nm. The combined beam was then expanded using a telescope with lenses $f_1 =$100 mm and $f_2 =$150 mm to improve the spatial mode matching at the input of the Ge-PCF. Finally, the signal and QD-band pump fields were combined with the C-band pump using a longpass dichroic mirror with a cut-on wavelength of 1180 nm. The combined beams were focused down to the Ge-PCF facet using an aspheric lens.

## 5. Results and discussion

### 5.1. Conversion within the QD-band and frequency comb generation

Table 1. The wavelengths used for conversion within the QD-band. The first three columns contain the signal and pump wavelengths used to generate the target wavelengths shown in the final column.

| $\lambda_s$ (nm) | $\lambda_p$ (nm) | $\lambda_q$ (nm) | $\lambda_t$ (nm) |
| --- | --- | --- | --- |
| 914.5 | 1555.7 | 1558.1–1560.9 | 912.6–913.5 |
| 917.2 | 1555.7 | 1558.1–1560.9 | 916.2–915.3 |
| 918.5 | 1555.7 | 1558.1–1560.9 | 917.5–916.5 |
| 921.5 | 1551.4–1554.8 | 1554.7 | 920.2–919.0 |

To explore the frequency conversion between various wavelengths around 920 nm, several datasets were gathered in which $\lambda_s$ and one of the pumps were kept constant, while the other pump was swept over a $\sim$ 3 nm range. Table 1 shows four combinations of wavelengths that we investigated, and the range of target wavelengths generated by propagation of those fields through the Ge-PCF. For each combination of $\lambda_s$, $\lambda_p$, and $\lambda_q$, in addition to the target wavelength which is blue-shifted from the signal, a red-shifted wavelength was also generated by the reverse interaction between the pumps (i.e. photon annihilation at $\lambda_q$ and creation at $\lambda_p$). Subsequently, many other frequencies were generated via cascaded mixing between the newly generated wavelengths and the pumps. These processes are also phase matched due to the near-degenerate regime in which we were working. As an example, the spectra generated for $\lambda_s$ = 917.2 nm and $\lambda_p$ = 1555.7 nm are plotted in Fig. 4(a). In a given spectrum, the conversion to each frequency is highly dependent on the coupling of the signal and pump fields into the Ge-PCF, as well as the polarization and pulse properties of the fields. Therefore, in every measurement, each of these variables was adjusted to optimize conversion to the target peak, which corresponds to the blue-shifted peak closest to $\lambda_s$. The spectra in (a) all demonstrate the depletion of the signal peak as photons are converted to other wavelengths, including that corresponding to the target peak. Fig. 4(b) zooms into the target peaks of the spectra from (a), showing how $\lambda_t$ was tuned across a nanometer range by tuning $\lambda_q$.

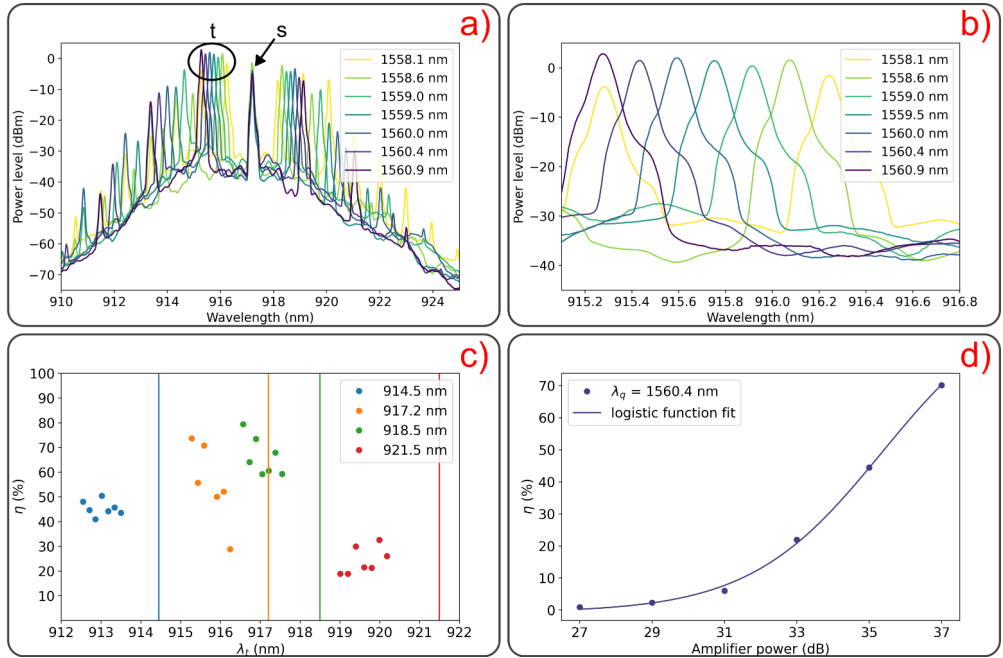

Fig. 4. (a) Spectra generated by propagation of the signal (s) and pump (p and q) fields through 34 m of Ge-PCF. In this case, $\lambda_s$ =917.2 nm, $\lambda_q$ =1558.1–1560.9 nm and $\lambda_p$ =1555.7 nm. Each curve corresponds to a different $\lambda_q$. The spectra were measured with a resolution of 0.05 nm (b) A zoomed-in view of the target (*t*) peaks encircled in (a), highlighting the tunability of the conversion. (c) Maximum conversion efficiency, $\eta$, as a function of the target wavelength. The coloured vertical lines correspond to the four signal wavelengths in the legend. $\eta$ was calculated as shown in Eq. 4. (d) Circles: $\eta$ from $\lambda_s$ =918.5 nm to $\lambda_t$ =916.7 nm as a function of amplifier power in dB. Where $\lambda_q$ =1560.4 nm and $\lambda_p$ =1555.7 nm. Line: logistic function fit.

The conversion efficiency to $\lambda_t$, $\eta$, was calculated using

$$\eta = 10^{\frac{\Delta P}{10}} \times 100\%, \quad (4)$$

where $\Delta P$ is the difference in the power level, in dB, of the target peak and the reference signal when the C-band beam is blocked. We assume near-uniform conversion efficiency across the pulse duration, as the C-band pumps are longer than the signal pulse. $\eta$ is plotted as a function of $\lambda_t$ in Fig. 4(c). Target wavelengths between 912.6 nm – 920.2 nm were generated with efficiencies ranging between 18.9% – 79.4%. To test whether the limiting factor in improving $\eta$ is conversion to other frequencies, $\eta$ is plotted as a function of amplifier power for $\lambda_s$ = 918.5 nm, $\lambda_t$ = 916.7 nm, $\lambda_q$ = 1560.4 nm, and $\lambda_p$ = 1555.7 nm in Fig. 4(d). If conversion to other frequencies is the main limitation, the rate at which $\eta$ grows would decrease with increasing amplifier power. The figure shows that the opposite is true, signifying that $\eta$ was instead limited by the power of the pumps, and even higher conversion efficiencies might be attainable with higher peak pump powers.

For these four sets of wavelength combinations, we found that conversion to $\lambda_t$ was generally higher than to the other generated wavelengths. However, by reducing the frequency spacing of the two pumps to a small fraction of the phase matching bandwidth around 918 nm, we also used our Ge-PCF to generate an optical frequency comb consisting of many peaks of comparable intensity. Frequency combs have many applications in optical communications and as sources of ultrafast pulses. Cascaded FWM in nonlinear fibers is well known as a successful approach to generating frequency combs and has been explored in other works [3, 26]. These demonstrations typically use three equally spaced pumps in the C-band, generating new frequencies nearby. In our example, one of the three pumps is instead well-separated in wavelength from the other two. This allows the parametric generation of a frequency comb at almost arbitrary wavelength (fixed by the group velocity matching condition chosen during the PCF design) from a single field at that wavelength. Results from our demonstration generating a frequency comb around 918 nm are shown in Fig. 5(a).

While frequency comb generation is an interesting application of our near-degenerate scheme, it should be suppressed in order to boost the efficiency $\eta$ of conversion to a single preferred target wavelength. We avoided significant conversion to frequencies other than our target empirically by sweeping our pump wavelengths and tuning the pump polarization until favourable conversion to $\lambda_t$ was observed. Further suppression of conversion to other frequencies could be achieved by narrowing the phase matching bandwidth by using a longer length of Ge-PCF than the 34 m used in the experiment. However, propagating the signal and pump fields through longer lengths of PCF would eventually introduce trade-offs between pulse power, duration and mode overlap due to the increasing effects of nonlinearity and dispersion. There is therefore a balance to be struck between narrowing the phase matching bandwidth to improve $\eta$, and inhibiting other nonlinear effects that can weaken $\eta$.

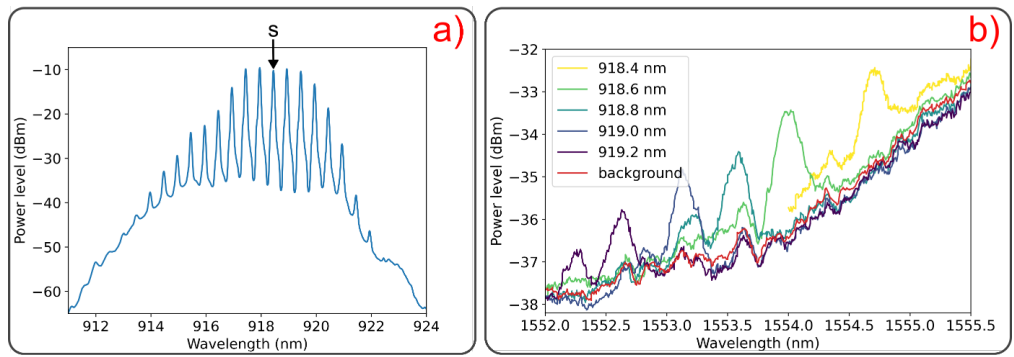

Fig. 5. (a) A frequency comb generated by propagation of $\lambda_s$ =918.4 nm, $\lambda_q$ = 1555.7 nm and $\lambda_p$ = 1554.2 nm through 34 m of Ge-PCF. (b) A zoomed-in view of target wavelengths generated by down-conversion from InAs QD wavelengths to the telecoms C-band, where $\lambda_s$ = 918.4–919.2 nm, $\lambda_q$ =1557.4 nm, and $\lambda_p$ =917.3 nm.

## 5.2. Downconversion of QD-band pulses to the C-band

In our second experiment targeting downconversion of QD-band pulses to wavelengths within the C-band, we fixed the wavelength of the EDFA, $\lambda_q = 1557.4$ nm, and the Ti:Sapphire laser, $\lambda_s = 917.3$ nm, while sweeping the M$^2$ laser, $\lambda_p$, over approximately 1 nm. In contrast to our earlier experiment where the pumps were closely spaced in frequency to enable small wavelength shifts within the QD emission band, here the frequency spacing of the pumps is larger, allowing a shift from the QD-band down to the C-band. The wavelength scheme for this process is shown in Fig. 3. Table 2 shows the range of $\lambda_s$ wavelengths investigated, and the range of target wavelengths subsequently generated in the C-band by propagation of each combination of the $s$, $q$, and $p$ fields through 34 m of the Ge-PCF. Fig. 5(b) plots these target peaks.

In this arrangement, the Ti:Sapphire laser was used to emulate InAs dot emission, with the limitation that the resulting down-conversion occurs within the classical regime. The efficiency of the frequency conversion was highly sensitive to several factors, such as the effective coupling of the input fields into the Ge-PCF, which maximizes optical power transfer, crucial for high conversion efficiency. In our setup, we achieved roughly 20% coupling efficiency for each of the input fields. Additionally, the polarization states of the input fields were optimized for maximum efficiency of frequency conversion. The pulse properties of the input fields were also critical. Specifically, the Ti:Sapphire pulses had a duration of approximately 60 ps and operate at a repetition rate of 76 MHz, resulting in a duty cycle of 0.46%. Since the M$^2$ is a continuous wave (c.w) laser, the temporal mode overlap efficiency corresponds to this duty cycle.

Due to the low temporal overlap between the the pulsed $q$ and $s$ fields and the c.w. $p$ field, the actual efficiency of downconversion is low. We can calculate a value of $\eta$ using Eq. 4 by taking $\Delta P$ to be the difference in the power level of the generated target peak (with the background removed and set on the -38.5 dBm measurement noise floor) and the reference $\lambda_s$ signal when the pump beams are blocked. The conversion efficiency of the system ranges between 0.06% – 0.09%, in line with expectations given the low peak power (~20 mW) of the c.w pump $p$. This demonstrates downconversion as a proof of principle using an unoptimized experimental setup, with conversion efficiencies approaching unity expected to be forthcoming with a more suitable choice of pump laser.

Table 2. The wavelengths used for downconversion of pulses from the QD-band to the C-band. The first three columns contain the signal and pump wavelengths used to generate the target wavelengths shown in the final column.

| $\lambda_s$ (nm) | $\lambda_p$ (nm) | $\lambda_q$ (nm) | $\lambda_t$ (nm) |
|---|---|---|---|
| 918.4–919.2 | 917.3 | 1557.4 | 1554.7–1552.6 |

## 6. Conclusions and outlook

We have demonstrated a near-degenerate Bragg-scattering FWM conversion scheme that achieves frequency conversion using bespoke Ge-PCF as a third-order nonlinear platform. The use of dopant in the core allows group velocity matching between the InAs emission band and the telecoms C-band without compromising confinement at long wavelengths. We demonstate the applicability of our fiber in three applications. First, it enables tunable conversion over a few-nm wavelength range to and from wavelengths around 920 nm with up to 79.4% internal conversion efficiency, which could be used to interface dissimilar QDs. Second, we showed how cascading this frequency conversion can be used to generate a frequency comb. Finally, it provides a platform for tunable conversion between the operation range of InAs QDs (~920 nm) and telecommunications wavelengths (~1550 nm). Future work will focus on demonstrating frequency conversion of single photons originating from InAs QDs to nearby wavelengths and the C-band while preserving their nonclassical statistics. This will be achieved by continued improvements in pump mode-matching, optical coupling, and noise filtering, paving the way to a fully fiber-integrated quantum frequency interfacing module.

**Funding.** LRM and AOCD acknowledge funding from UKRI EPSRC grant "PHOCIS: A Photonic Crystal Integrated Squeezer" EP/W028336/1. PA, AJB, PJM and AOCD acknowledge support from the UK Hub in Quantum Computing and Simulation, part of the UK National Quantum Technologies Programme with funding from UKRI EPSRC (EP/T001062/1). AJB acknowledges financial support provided by EPSRC via Grant No. EP/T017813/1. MA was supported by grant EP/S024441/1 and the National Physical Laboratory.

**Acknowledgments.** We would like to thank Steven Renshaw for assisting with the fiber fabrication.

**Disclosures.** The authors declare no conflicts of interest.

**Data Availability Statement.** Data underlying the results presented in this paper are available in [27].